\documentclass[aps,prc,twocolumn,superscriptaddress,nofootinbib,showpacs]{revtex4-1}

\usepackage{amsmath,amssymb,amscd}
\usepackage{listings}
\usepackage{dsfont}
\usepackage{slashed}
\usepackage{color}

\usepackage{graphicx}
\usepackage{epstopdf}
\usepackage{subfigure}
\usepackage{epsfig}
\usepackage[breaklinks,hidelinks]{hyperref}
\usepackage{latexsym}
\usepackage{multirow}
\usepackage{mciteplus}

\newcommand{\bea}{\begin{eqnarray}}
\newcommand{\eea}{\end{eqnarray}}
\newcommand{\be}{\begin{equation}}
\newcommand{\ee}{\end{equation}}
\newcommand{\beq}{\begin{equation}}
\newcommand{\enq}{\end{equation}}

\newcommand{\uN}{\mathcal{N}}

\newcommand{\uS}{\mathcal{S}}

\def\astat{\alpha_{E1}}
\def\bstat{\beta_{M1}}
\def\edip{\alpha_{E1}(\omega)}
\def\mdip{\beta_{M1}(\omega)}
\def\ediplex{\alpha_{E1}^{\text{L}}(\omega)}
\def\mdiplex{\beta_{M1}^{\text{L}}(\omega)}
\def\ediplexzero{\alpha_{E1}^{\text{L}_0}(\omega)}
\def\mdiplexzero{\beta_{M1}^{\text{L}_0}(\omega)}

\newcommand{\benum}{\begin{enumerate}}
\newcommand{\enum}{\end{enumerate}}
\newcommand{\beitem}{\begin{itemize}}
\newcommand{\enitem}{\end{itemize}}
\newcommand{\eqr}[1]{Eq.~(\ref{#1})}
\newcommand{\eqrtwo}[2]{Eqs.~(\ref{#1}) and (\ref{#2})}

\newcommand{\tref}[1]{Table~\ref{#1}}

\newcommand{\fr}[1]{Fig.~\ref{#1}}

\newcommand{\cref}[1]{Ch.~\ref{#1}}

\newlength\savedwidth

\usepackage[normalem]{ulem}
\renewcommand\sout{\bgroup \color[rgb]{0.55,0.00,0.99} \ULdepth=-.5ex \ULset}

\draft

\allowdisplaybreaks
\begin{document}

\author{B.~Pasquini}

\affiliation{Dipartimento di Fisica,
Universit\`a degli Studi di Pavia, 27100 Pavia, Italy}
\affiliation{Istituto Nazionale di Fisica Nucleare,
Sezione di Pavia, 27100 Pavia, Italy}

\author{P. Pedroni}
\affiliation{Istituto Nazionale di Fisica Nucleare,
Sezione di Pavia, 27100 Pavia, Italy}

\author{S. Sconfietti}

\affiliation{Dipartimento di Fisica,
Universit\`a degli Studi di Pavia, 27100 Pavia, Italy}
\affiliation{Istituto Nazionale di Fisica Nucleare,
Sezione di Pavia, 27100 Pavia, Italy}
\title{First extraction of the scalar proton dynamical polarizabilities from real Compton scattering data}

\begin{abstract} 
We present the first attempt to extract the  scalar dipole dynamical polarizabilities from proton real Compton scattering data below pion-production threshold. 
The theoretical framework combines dispersion relations technique, low-energy expansion and multipole decomposition of the scattering amplitudes.
The results are obtained with statistical tools that have never been applied so far to  Compton scattering data and are crucial to overcome problems inherent
to the analysis of the available data set. 
\end{abstract} 
\maketitle 

\section{Introduction}
\label{sec:intro}

Real Compton scattering (RCS) is one of the fundamental processes to access information on the internal structure of the nucleon. 
The RCS amplitude can be separated into a Born contribution, describing the scattering off a pointlike nucleon with anomalous magnetic moment, and a structure-dependent part
referred as non-Born term.
The non-Born contribution is parametrized by polarizabilities, which describe the  response of the  nucleon's internal degrees of freedom 
 to an external electromagnetic field. 
 In the low-energy expansion of the non-Born amplitudes, the leading-order effects are given by static polarizabilities, that are defined  in the limit of zero frequency of the photon field
 and therefore measure the response to a static external electromagnetic field.
 The leading-order spin-independent polarizabilities are the scalar dipole electric and magnetic polarizabilities, $\alpha_{E1}$ and $\beta_{M1}$, respectively,  while four spin-dependent 
 polarizabilities appear at the next order and involve  the nucleon-spin degrees of freedom.
They have been the subject of intense research both experimentally and theoretically~\cite{Babusci:1998ww,Drechsel:2002ar,Schumacher:2005an,Griesshammer:2012we,Hagelstein:2015egb,Pasquini:2018wbl}. The currently accepted values for the proton scalar polarizabilities are
 $\alpha_{E1}=(11.2\pm0.4)\cdot10^{-4}\text{fm}^3$
and $\beta_{M1}=(2.5\mp0.4)\cdot10^{-4}\text{fm}^3$~\cite{Patrignani:2016xqp}, while
the first extraction of the individual four spin polarizabilities has been obtained only recently from double-polarized Compton scattering~\cite{Martel:2014pba}. 
\\
As it is well known from many branches of physics, polarizabilities  become energy
dependent due to internal relaxation mechanisms, resonances and particle production
thresholds in a physical system~\cite{0143-0807-29-3-010,Kittel,1977JChPh..66..191G}. This energy dependence is subsumed in the definition of dynamical polarizabilities, that 
 parametrize the response of the internal degrees of freedom of a composite object to an external,
real-photon field of arbitrary energy. 
  The enriched information encoded in the dynamical nucleon polarizabilities has been pointed out in different theoretical calculations, using dispersion relations or effective field theories~\cite{Griesshammer:2001uw,Hildebrandt:2003fm,Aleksejevs:2013cda,Aleksejevs:2010zw,Lensky:2015awa}.
In this work, we attempt for the first time to extract information on the scalar dipole  dynamical polarizabilities (DDPs) from a fit to all available unpolarized RCS data below pion-production threshold. To this aim, we apply a  statistical analysis  based on the {\em parametric bootstrap} technique (see, for instance,~\cite{Davidson-Hinkley} and  references therein).
  Such a method has never been exploited to analyze RCS data and it is crucial to deal with problems inherent to both  the low sensitivity of the RCS cross section to the energy dependence of the  dynamical polarizabilities and to the poor accuracy of the available data sets.

A feasibility study  for the extraction  of two  spin  dynamical polarizabilities from unpolarized Compton scattering data has been presented previously in Refs.~\cite{Hildebrandt:2005ix,Griesshammer:2004yn}, following a different strategy from the present work. 
These fits
 did not turn out to be conclusive, mainly because of  the scarce accuracy of the data set at  disposal and the smaller sensitivity of the unpolarized cross section to the spin polarizabilities rather than the scalar polarizabilities. More recently, a partial-wave analysis of the unpolarized Compton scattering data has been discussed also in Ref.~\cite{Krupina:2017pgr}, pointing out the need to improve the accuracy of the experimental data set to pin down the values of the static scalar dipole polarizabilities with more precision.

  The paper is organized as follows: In Sec.~\ref{sec:theory}, we introduce the theoretical framework to analyze the unpolarized Compton scattering cross section.
  In Sec.~\ref{sec:strategy} we  describe and motivate our fitting procedure based on the bootstrap technique, in comparison with the standard chi-squared minimization technique.
  Section~\ref{sec:results} contains   our results for the fit of the scalar DDPs, and section~\ref{sec:conclusion} summarizes our conclusions.

  \section{Theoretical framework}
\label{sec:theory}
The theoretical framework for the analysis relies on the multipole expansion of the scattering amplitude, the low-energy expansion (LEX) of the scalar DDPs and dispersion relations (DRs) for the calculation of the higher-order multipole amplitudes and  the energy dependence of the scalar DDPs near the pion-production threshold.
\\
The definition of the dynamical polarizabilities rests on the multipole expansion of the RCS amplitude in the center-of-mass (cm) frame~\cite{Babusci:1998ww,Lapidus:1960,Ritus,Contogouris1962}.
The multipole amplitudes $f^{l\pm}_{TT'}$
with $T,$  $T' = E,M$, correspond to transitions $Tl \rightarrow T'l'$  and the superscript indicates the
angular momentum $l$ of the initial photon and the total angular momentum $j = l\pm1/2$.
 In particular, for the scalar DDPs one has~\cite{Hildebrandt:2003fm}:
	\bea
	\label{eq:ddp}
\edip = \frac{2\bar{f}_{EE}^{1+} + \bar{f}_{EE}^{1-}}{\omega^2}, \,
 \mdip &=& \frac{2\bar{f}_{MM}^{1+} + \bar{f}_{MM}^{1-}}{\omega^2},
	\label{eq:ddp-beta}
	\eea
	where $\bar{f}$  indicates the non-Born contribution to the multipoles.
	 In  the calculation of the cross section, we  compute the full Born contribution, 
	given by pole diagrams involving a single nucleon exchanged
in $s$- or $u$-channels and $\gamma NN$ vertices taken in the on-shell regime~\cite{Babusci:1998ww}.
For the non-Born part, we use the multipole expansion.
As also observed  in Ref.~\cite{Hildebrandt:2003fm}, the multipole expansion of the non-Born contribution has a very fast convergence below pion-production threshold. In our analysis, we take into account the non-Born amplitudes  $\bar{f}^{l\pm}_{TT'}$ up to $l=3$: the scalar DDPs are fitted to the data, and the remaining contributions are calculated through subtracted DRs.

DRs have been proven to be a powerful tool to analyze RCS data~\cite{Schumacher:2005an,Lvov:1996rmi,Pasquini:2010zr,Martel:2014pba,OlmosdeLeon:2001zn}, as they allow to minimize the model dependence using as input available experimental information from other processes.
The dispersion calculation is performed in terms of six independent invariant amplitudes $A_i$, $i=1,\dots,\, 6$, which can be recast in terms of the multipole amplitudes  $f^{l\pm}_{TT'}$ as explained in App. A of Ref.~\cite{Hildebrandt:2003fm}.
In the subtracted formalism, 
the six invariant amplitudes are obtained from subtracted dispersion integrals in both the $s$  and $t$ channels, and subtraction constants that are directly related to the six leading-order static polarizabilities.
The subtracted integrals are saturated by $\pi N$, $\pi \pi N$ and heavier-meson intermediate states in the $s$ channel, and $\pi \pi$ intermediate states in the $t$ channel~\cite{Drechsel:1999rf}.
In the present analysis, the input for the pion-photoproduction amplitudes has been updated to the most recent version of MAID~\cite{Drechsel:2007if}, while we refer to~\cite{Drechsel:1999rf,Pasquini:2007hf}  for the calculation of the contributions beyond $\pi N$  in the $s$-channel and for the $t$-channel contribution.
Four of the subtraction constants are fixed to the values of  the  static leading-order spin polarizabilities extracted in Ref.~\cite{Martel:2014pba}. The two remaining  constants
are given in terms of the static dipole scalar polarizabilities as specified in the following.

We are interested in  the energy dependence of the scalar DDPs below pion-production threshold, where they are real functions. 
By performing a LEX of the scalar DDPs, one recovers the limiting values 
 of the static dipole polarizabilities  at zero energy.
 Higher-order terms contain dispersive or retardation effects which can be parametrized in terms of higher-order polarizabilities~\cite{Babusci:1998ww,Holstein:1999uu,Lensky:2015awa}.
 
 The static polarizabilities are best defined via the effective non-relativistic Hamiltonian in the Breit
frame, and  hence the next-to-leading-order coefficients of the LEX of Eq.~\eqref{eq:ddp}  are not  given  only in terms of  higher-order static polarizabilities related to retardation effects of the $E1$ and $M1$ radiation.
The direct link is spoiled by recoil corrections in the cm frame.
 We have calculated  the relevant recoil corrections  up to ${\cal O}(\omega^5)$, obtaining  the following expressions for the LEX of the scalar DDPs:
\begin{widetext}
\begin{eqnarray}
\ediplex&=&\alpha_{E1}+\frac{\beta_{M1}}{M}\omega+\left(\alpha_{E1,\nu}+\frac{5 \alpha_{E1}-2 \beta_{M1}}{8 m^2}\right)\omega^2 +\left(\frac{8 \alpha_{E1,\nu}+\alpha_{E2}+12
   \beta_{M1,\nu}}{8 M} +\frac{\gamma_{M1E2}-\gamma_{M1M1}}{8 M^2} \right.\nonumber\\
&+&\left.\frac{\beta_{M1}-2 \alpha_{E1}}{8 M^3}\right)\omega^3+\left\{\alpha_4^{\text{L}}
-\frac{1}{40M}[15( \gamma_{E1E1,\nu}- \gamma_{E1M2,\nu})-69
   \gamma_{E2E2}+12 (\gamma_{E2M3}-\gamma_{M2E3})\right.\nonumber\\
   &+&25(\gamma_{M1E2,\nu}-\gamma_{M1M1,\nu})+51
   \gamma_{M2M2}]+\frac{1}{480 M^2} (1248 \alpha_{E1,\nu}+95 \alpha_{E2}+540 \beta_{M1,\nu}+26
   \beta_{M2})\nonumber\\
&
+&\frac{1}{80 M^3} [25( \gamma_{E1E1}- \gamma_{E1M2})+39(\gamma_{M1E2}-\gamma_{M1M1})]-\left.\frac{1}{160 M^4}(24\alpha_{E1}+19\beta_{M1})\right\}\omega^4\nonumber\\
&+& \left\{\alpha_5^{\text{L}}+
\frac{1}{200M^2}  (55 (\gamma_{E1E1,\nu}-\gamma_{E1M2,\nu})-6(35\gamma_{E2E2}-22\gamma_{E2M3}+5\gamma_{M1E2,\nu}\right.\nonumber\\
&-&5\gamma_{M1M1,\nu}+38\gamma_{M2E3})+555
   \gamma_{M2M2})+\frac{1}{480M^3} (612 \alpha_{E1,\nu}+38 \alpha_{E2}+1008 \beta_{M1,\nu}+89
   \beta_{M2})\nonumber \\
&+&\frac{1}{160M^4} (-46 (\gamma_{E1E1}- \gamma_{E1M2})+33
   (\gamma_{M1M1}-\gamma_{M1E2}))+\left.\frac{1}{160 M^5}(\alpha_{E1}-14\beta_{M1})\right\}\omega^5,
\label{alpha}
\\\mdiplex&=&\beta_{M1}+\frac{\alpha_{E1}}{M}\omega+\left(\beta_{M1,\nu}+\frac{5 \beta_{M1}-2 \alpha_{E1}}{8 M^2}\right)\omega^2+\left(\frac{8 \beta_{M1,\nu}+\beta_{M2}+12
   \alpha_{E1,\nu}}{8 M}+\frac{\gamma_{E1M2}-\gamma_{E1E1}}{8 M^2}\right.\nonumber\\
&+&\left.\frac{\alpha_{E1}-2 \beta_{M1}}{8 M^3}\right)\omega^3+\left\{\beta_4^{\text{L}}
-\frac{1}{40 M} [15 (\gamma_{M1M1,\nu}- \gamma_{M1E2,\nu})-69
   \gamma_{M2M2}+12 (\gamma_{M2E3}- \gamma_{E2M3})\right.\nonumber\\
&+&25 (\gamma_{E1M2,\nu}- \gamma_{E1E1,\nu})+51
   \gamma_{E2E2}]+\frac{1}{480M^2} (1248 \beta_{M1,\nu}+95 \beta_{M2}+540 \alpha_{E1,\nu}+26
   \alpha_{E2})\nonumber\\
&+&\frac{1}{80 M^3} [25( \gamma_{M1M1}- \gamma_{M1E2})+39(\gamma_{E1M2}-\gamma_{E1E1})]-\left.\frac{1}{160 M^4}(24 \beta_{M1}+19 \alpha_{E1})\right\}\omega^4\nonumber\\
&+& \left\{\beta_5^{\text{L}}+\frac{1}{200M^2} [55 (\gamma_{M1M1,\nu}- \gamma_{M1E2,\nu})-6(35
   \gamma_{M2M2}+22\gamma_{M2E3}-5 \gamma_{E1M2,\nu}\right.\nonumber\\
&+&5 \gamma_{E1E1,\nu}-38 \gamma_{E2M3})+555
   \gamma_{E2E2})+
\frac{1}{480 M^3}
 (612 \beta_{M1,\nu}+38 \beta_{M2}+1008 \alpha_{E1,\nu}+89
   \alpha_{E2})\nonumber \\&+&\frac{1}{160M^4}  (-46 (\gamma_{M1M1}- \gamma_{M1E2})+33
   (\gamma_{E1E1}-\gamma_{E1M2}))+\left.\frac{1}{160 M^5}( \beta_{M1}-14 \alpha_{E1})\right\}\omega^5.
\label{beta}
\end{eqnarray}
\end{widetext}
In Eqs.~\eqref{alpha} and~\eqref{beta}, the  terms with even power of $\omega$ contain both retardation effects of the dipole radiations and recoil terms. The terms with odd powers of $\omega$ are recoil contributions, which, in addition to the contributions from the static polarizabilities of lower orders, can include terms with the static scalar and spin-dependent polarizabilities of higher multipolarity. \\
In particular, the first dispersive contributions enter at ${\cal O}(\omega^2)$ and correspond to the static polarizabilities
$\alpha_{E1,\nu}$ and $\beta_{M1,\nu}$.
The recoil terms at ${\cal O}(\omega^3)$ are given in terms of the static dipole scalar and spin polarizabilities, the fourth-order dipole scalar polarizabilities $\alpha_{E1,\nu}$ and
$\beta_{M1,\nu}$  and the quadrupole scalar polarizabilities $\alpha_{E2}$  and $\beta_{M2}$. In particular,  $\alpha_{E1,\nu}$, $\beta_{M1,\nu}$ and the quadrupole polarizabilities enter as recoil terms with the same suppression factor in $1/M$. The static spin dipole polarizabilities enter with a coefficient in $1/M^2$ and the static  scalar dipole polarizabilities enter with a factor in $1/M^3$. 
\\ 
At ${\cal O}(\omega^4)$, the recoil terms contain different combinations of the same polarizabilities entering at ${\cal O}(\omega^3)$, weighed with an additional power in $1/M$.  Furthermore, they involve the higher-order spin polarizabilities defined in Ref.~\cite{Holstein:1999uu}, with a coefficient in $1/M$,
 and the dispersive coefficients $\alpha_4^{\text{L}}$ and $\beta_4^{\text{L}}$ 
 corresponding to sixth-order scalar polarizabilities, which have never been defined in literature.
Following Ref.~\cite{Babusci:1998ww}, we can write them as combinations of the second-order derivatives of the non-Born contribution to the Lorentz invariant amplitudes $A_i$, i.e.
\begin{eqnarray}
a_{i,\nu\nu}=\frac{\partial A_i^{\mathrm{NB}}}{\partial^2 \nu^2},\quad a_{i,\nu t}=\frac{\partial^2 A_i^{\mathrm{NB}}}{\partial \nu^2\partial t},\quad a_{i,t t}=\frac{\partial^2 A_i^{\mathrm{NB}}}{\partial^2 t}, \nonumber\\
( i=1,\dots, 6).\label{structure-constants}
\end{eqnarray}
At ${\cal O}(\omega^5)$, one finds  a recoil contribution in $1/M$ given by combinations of these 18 constants, which correspond to a combination of the dispersive effects of the sixth-order scalar polarizabilities entering at  ${\cal O}(\omega^4)$ and new scalar sixth-order polarizabilities, which have never been discussed so-far in literature.
These terms are collectively indicated with the $\alpha_5^{\text{L}}$ and $\beta_5^{\text{L}}$ coefficients  in Eqs.~\eqref{alpha} and \eqref{beta}, respectively.
\begin{figure}[t]
  \includegraphics[scale=.4]{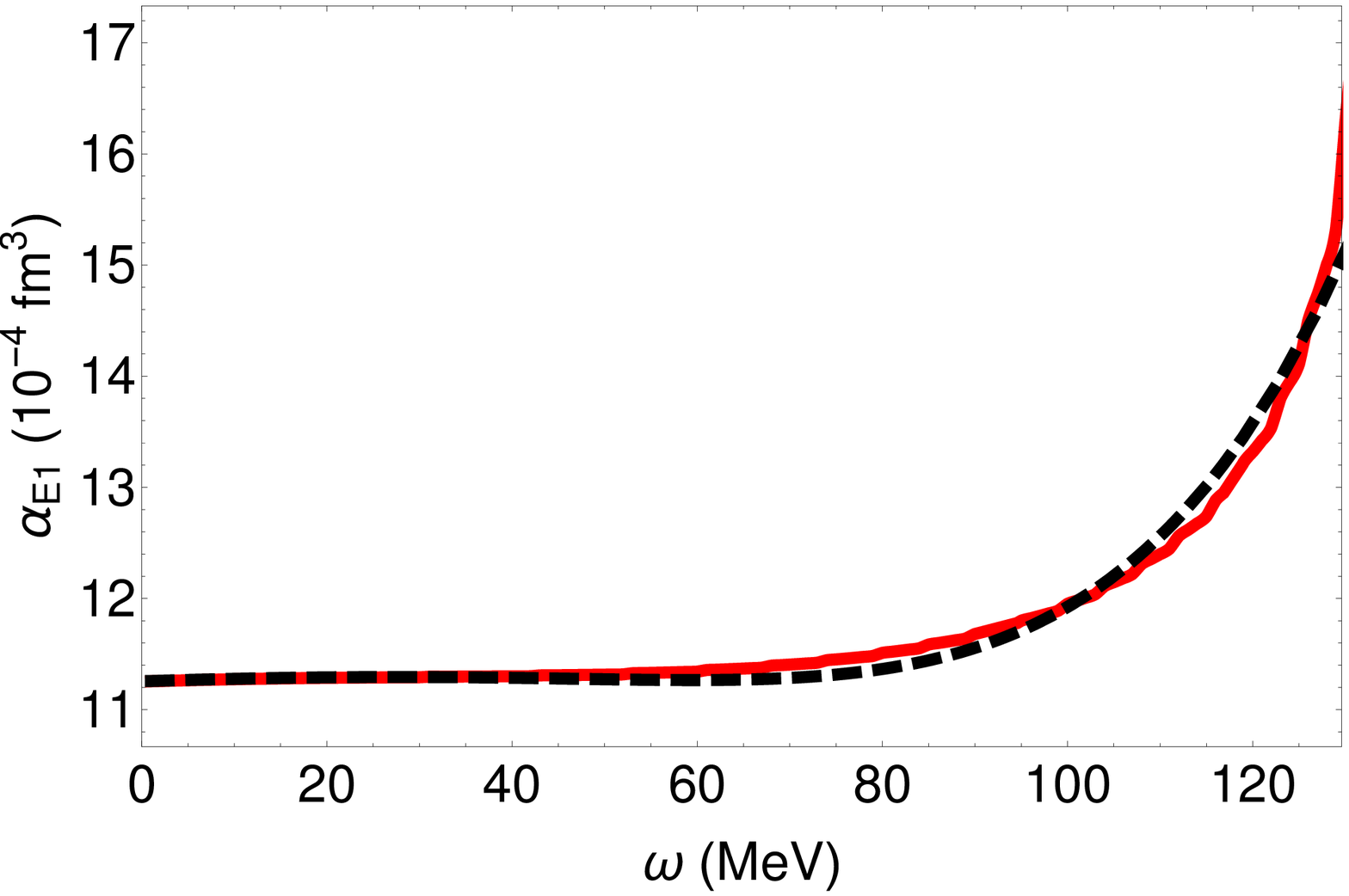}
  \includegraphics[scale=.4]{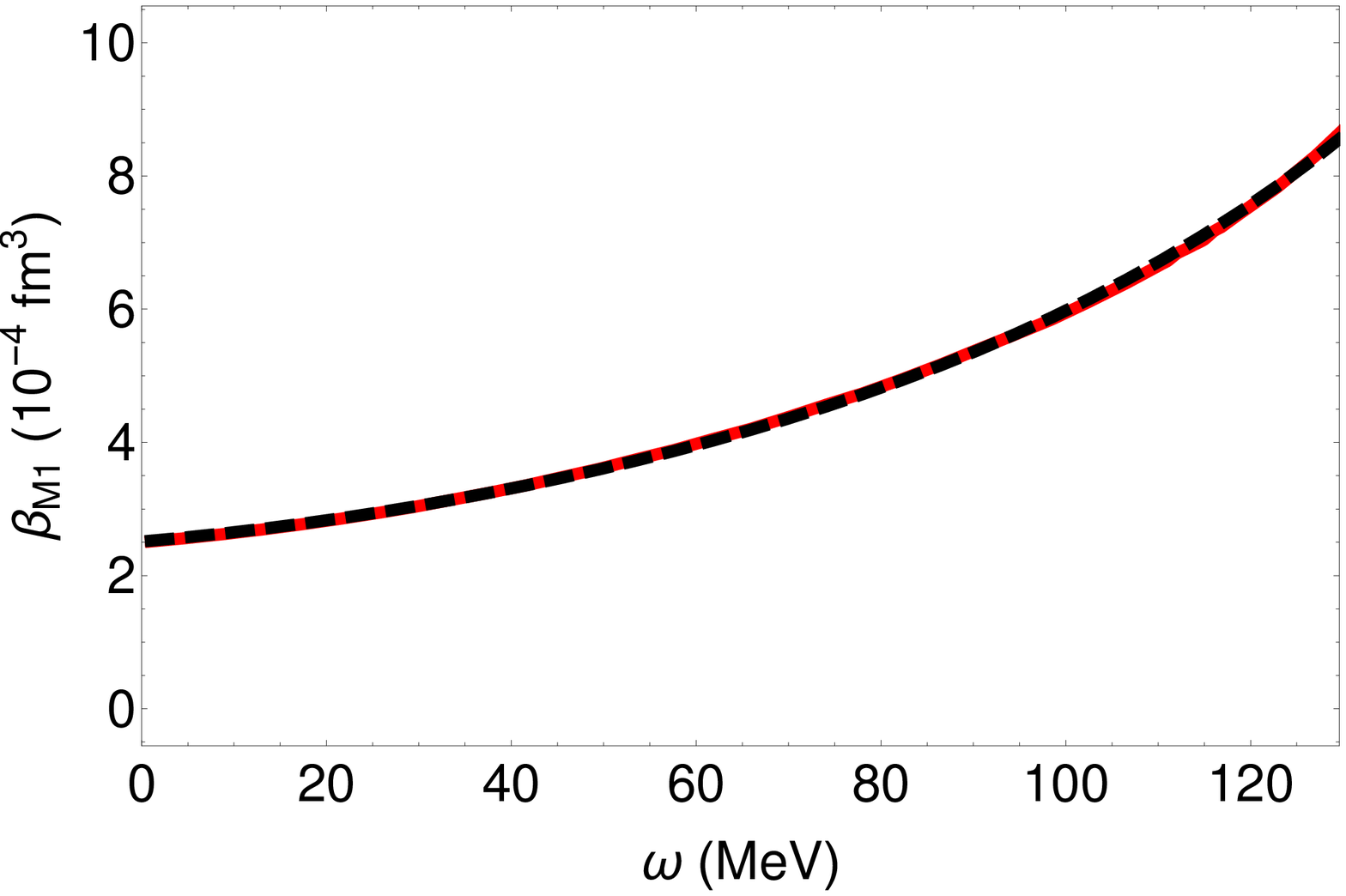}
  \caption{ (Color online) The scalar DDPs $\alpha_{E1}(\omega)$ (upper panel) and $\beta_{M1}(\omega)$ (lower panel) as function of the centre of mass energy $\omega$.
  The solid red curves show the results from the  DR calculation, with the full energy dependence.
  The dashed black curves are the results from Eq.~\eqref{eq:lex3},  using the predictions from DRs for all the static polarizabilities entering the $\ediplexzero$ and $\mdiplexzero$ contributions and  the results from the fit to the DR calculation  for the residual functions $f_\alpha(\omega)$ and $f_\beta(\omega)$.}
  \label{fig:alpha-beta}
	\end{figure}
\\
The convergence radius of such Taylor expansion is limited by the first singularity, which is set by the pion-production branch cut. 
In particular, the LEX of  $\alpha_{E1}(\omega)$ fails to reproduce the non-analytical behaviour of the polarizability when approaching the pion production threshold.
The contribution beyond the 
 LEX in Eqs.~\eqref{alpha} and \eqref{beta} can be taken into account by introducing  two residual functions $\tilde f$ defined by
	\beq
\edip = \ediplex + \tilde f_\alpha(\omega), \quad \mdip = \mdiplex +\tilde f_\beta(\omega).
\label{eq:lex2}
	\enq
The two functions $\tilde f_\alpha(\omega)$ and $\tilde f_\beta(\omega)$ can be calculated using DRs, and the results from DRs can be parametrized
using the following functional form 
$\tilde f_\alpha(\omega)=\tilde\alpha_4\omega^4+\tilde\alpha_5\omega^5$ and $\tilde f_\beta(\omega)=\tilde\beta_4\omega^4+\tilde\beta_5\omega^5$.
This particular choice allows us to merge the $\alpha_4^{\text{L}},$ $\alpha_5^{\text{L}},$ $\beta_4^{\text{L}}$ and $\beta_5^{\text{LL}}$ coefficients 
in  Eqs.~\eqref{alpha} and \eqref{beta} 
with the polynomial coefficients of   $\tilde f_{\alpha,\,\beta}(\omega)$ and to write the whole energy dependence
of the scalar DDPs  as
	\beq
\edip = \ediplexzero +  f_\alpha(\omega), \quad \mdip = \mdiplexzero + f_\beta(\omega),
\label{eq:lex3}
	\enq
where
\bea
\label{eq:lex0}
\, \ediplexzero \equiv \ediplex|_{\alpha_4^{\text{L}}=\alpha_5^{\text{L}}=0}, \quad  f_\alpha(\omega) \equiv \alpha_4\omega^4 + \alpha_5\omega^5,\nonumber \\
\mdiplexzero \equiv \mdiplex|_{\beta_4^{\text{L}}=\beta_5^{\text{L}}=0}, \quad f_\beta(\omega) \equiv \beta_4\omega^4 + \beta_5\omega^5.\nonumber\\
	\eea
	with
	\bea
	\alpha_{4,5} \equiv \alpha_{4,5}^{\text{L}} + \tilde\alpha_{4,5},\quad
	\beta_{4,5}\equiv \beta_{4,5}^{\text{L}}+,\tilde\beta_{4,5}.\label{coefficients-fit}
	\eea
The analytical expressions for the scalar DDPs in Eqs.~\eqref{eq:lex3}  is the same as in \eqrtwo{alpha}{beta}, provided that the $(\alpha,\beta)_{4,5}^{\text{L}}$ coefficients are replaced by the coefficients $(\alpha,\beta)_{4,5}$ in Eqs.~\eqref{coefficients-fit}.

In Fig.~\ref{fig:alpha-beta}, we show the predictions for the scalar DDPs from the full DR calculation, without LEX, 
in comparison with the results obtained from Eqs.~\eqref{eq:lex3}, using the predictions from DRs for all the static polarizabilities entering the $\ediplexzero$ and $\mdiplexzero$ contributions and  the results from the fit to the DR calculation  for the residual functions $f_\alpha(\omega)$ and $f_\beta(\omega)$.
We note that the parametrization in Eq.~\eqref{eq:lex3} is able to reproduce very well the full energy dependence of the scalar DDPs in the energy range considered in the present fit, 
 giving us confidence that it can be conveniently adopted for our fitting procedure of the scalar DDPs to the Compton scattering data.

\section{Fitting strategy}
\label{sec:strategy}
In this section, we outline
the fitting strategy for the coefficients of the LEX in \eqr{eq:lex3}.
As we are interested into the genuine dispersive effects of the $E1$ and $M1$ radiation in the LEX of the scalar DDPs, we fixed  the recoil contributions in $\ediplexzero$ and $\mdiplexzero$ from the static
leading-order spin polarizabilities to the experimental values of Ref.~\cite{Martel:2014pba}, and  the recoil terms from higher-order spin polarizabilities as well as from  quadrupole scalar polarizabilities to the values predicted  from
subtracted DRs~\cite{Holstein:1999uu}. 

As input for the subtracted dispersion integrals we used the updated MAID solution, which is employed also for the calculation of the multipole amplitudes which are not fitted.
We used two different data sets, i.e., all the available experimental data for the unpolarized cross sections below pion-production threshold, 
denoted as "FULL" data set\footnote{For the data sets of \cite{Oxley:1958zz,Hyman:1959zz,GOLDANSKY1960473} and \cite{Pugh:1957zz} we used the compilation of Baranov \cite{Baranov:2001jv}, as also done in Ref.~\cite{Griesshammer:2012we}.}  (for a total of 150 data points)~\cite{OlmosdeLeon:2001zn,Federspiel:1991yd,Zieger:1992jq,MacGibbon:1995in,GOLDANSKY1960473,Hallin:1993ft,Hyman:1959zz,Pugh:1957zz,Bernardini1960,Baranov:1974ec,Baranov:1975ju,Oxley:1958zz}, and  the  data set given by the TAPS experiment alone 
(55 data points)~\cite{OlmosdeLeon:2001zn}, which is, by far, the most comprehensive available subset. ``Improved" data sets have been  defined in recent fits of RCS observables, by discarding some data points from different experiments~\cite{Griesshammer:2012we,McGovern:2012ew,Krupina:2017pgr}. 
Here we do not apply any selection  to the data, and we postpone a more detailed discussion of the statistical consistency of the different data subsets to a future work~\cite{preparation}, entirely devoted to the extraction of the static scalar dipole polarizabilities from data. \\
As a first attempt, we tried to fit $\astat, $ $\bstat,$ $ \alpha_{E1,\nu},$  $\beta_{M1,\nu}$ and
 the four coefficients parametrizing the residual functions $f_\alpha(\omega)$ and $f_\beta(\omega)$, for a total of eight parameters.
This choice has the  advantage of fitting the full energy dependence of the scalar DDPs with a minimum of model dependence.
We then applied the  gradient method in the $\chi^2$ minimization procedure of MINUIT~\cite{James:1975dr}. 
Unfortunately, this method did not show convergence since the  positive-definiteness condition of the covariance matrix could not be achieved.
This is because of too strong correlations between the fitted parameters, resulting from the very low 
sensitivity of the available experimental data to the higher order dispersive coefficients
(note that the fitting parameters enter to all orders  in the LEX of the scalar  DDPs, both as genuine dispersive effects and as recoil contributions).

\noindent 
To circumvent this problem, we used a
combination of the 
{\em simplex}~\cite{Nelder:1965zz} (that is a purely geometric minimization algorithm) 
and bootstrap (that is a Monte Carlo technique) methods. 
Each bootstrap ``measurement" is assumed to be Gaussian distributed around a given experimental data point 
 with a standard deviation given by its statistical error. 
 All bootstrapped points of a given subset are then shifted by the same quantity proportional to the published systematic error, 
 assumed to be uniformly distributed.
If we define as {\em cycle} a number of bootstrapped points equal to the total number of points 
in the considered experimental data set, the bootstrap sampling can be finally described as: 
	\beq
	\label{eq:boot}
\uS_{i,k,j} = \xi_{k,j}\left[\uS_i^{\text{exp}} {+} \gamma_{i,j}\sigma_i^{\text{exp}}\right],
	\enq
where $\uS$ stands for the differential cross section, with the superscript ``$\text{exp}$" indicating the experimental values for the mean value $\uS_i^{\text{exp}}$ and the statistical error $\sigma_i^{\text{exp}}$. 
In \eqr{eq:boot}, the index $i$ runs over the data points in the whole set, the index $k$ labels each subset, and the index $j$ indicates the bootstrap cycle. Furthermore, $\xi_{k,j}$  are random numbers uniformly distributed as $\mathcal U[1-\Delta_k,1+\Delta_k]$ 
(with {$\pm \Delta_k$} the published systematic error), while the numbers $\gamma_{i,j}$ are sampled from the standard Gaussian distribution $\uN[0,1]$.
When different systematic-error sources are quantified, $\xi_{k,j}$ is the combination of all the
contributions.
According to Ref.~\cite{OlmosdeLeon:2001zn},  $\sigma_i^{\text{exp}}$ for the TAPS data set  includes a $\pm$5\% point-to-point systematic error added in quadrature to the statistical error of each individual data point.

The minimization is performed after a complete cycle, and the output for the fitted values of the polarizabilities is stored. 
Repeating  the bootstrap cycle a very large $n_R$ number of replicas (we choose $n_R=10000$), we are finally able to reconstruct 
the probability distributions for every fitted parameter.
\\
In order to obtain a cross-check for our fitting method, we first assumed as fit parameters only $\astat-\bstat$,
using the constraint from the Baldin's sum rule for the polarizability sum, with the TAPS value  $\astat+\bstat=(13.8\pm0.4)\cdot 10^{-4}$ fm$^3$~\cite{OlmosdeLeon:2001zn}\footnote{{ This value is consistent  with the weighed average over the available evaluations of the Baldin sum rule~\cite{Hagelstein:2015egb}.}}, 
fixing the leading-order spin polarizabilities to the central  values of Ref.~\cite{Martel:2014pba}, all the other static
polarizabilities as well as the residual functions $f_\alpha(\omega)$ and $f_\beta(\omega)$  to subtracted  DRs.
This configuration with  only one free parameter allows the gradient method to converge, thus
providing a benchmark 
both for our new minimization algorithm and for the theoretical framework.
 The validation of the method should pass the following tests:
	\beitem
      \item The one-parameter fit with our strategy (the combination of subtracted DRs, LEX, multipole expansion and bootstrap technique, labeled as fit 1)
        should be consistent with the fit using the complete DR calculation (without  multipole expansion and  LEX), and the gradient method for the minimization of  the $\chi^2$ function (labeled as fit 2);
      \item The results of the fit 1 should be consistent with the results obtained from the combination of subtracted DRs, LEX, multipole expansion and gradient method
        (labeled as fit 3).
  \enitem
 For the purposes of this test, the bootstrap sampling procedure was performed after fixing to unity all $\xi_{k,j}$ parameters  in \eqr{eq:boot}.
 The results from the three fits using the FULL data set are shown in  \tref{tab:stat_check}. They are all consistent  with each other and
 also agree with the PDG values~\cite{Patrignani:2016xqp}.
   In addition, the errors evaluated using fit 1 are  
     Gaussian-distributed, in agreement with the statistical expectations (see, for instance,~\cite{James:2006zz}).

The bootstrap solution does not significantly change when 
 the uncertainty in the  Baldin's sum rule value is taken into account by randomly generating, for each cycle, a
 different $\astat+\bstat$ value sampled from the $\uN[13.8,(0.4)^2]$ distribution.
  Similarly, the results of the fit parameters change at most by $1\%$
   when  the values of the leading-order spin polarizabilities
   are varied within the uncertainties quoted in Ref.~\cite{Martel:2014pba}.

     All this  gives us confidence in both the theoretical framework, 
based on the LEX and multipole expansion, and the statistical tools based on the bootstrap technique. In summary, the main advantages of the adopted technique are:
\beitem 
\item The straightforward inclusion of systematic errors in the minimization procedure, as shown in \eqr{eq:boot}. 
This feature allows us to reduce the overall number of fit parameters with respect to the 
extended-$\chi^2$ procedure,  with a normalization factor for each data set left as free parameters~\cite{Griesshammer:2012we};
\item The fact that any error distribution of the experimental data can be easily implemented. Moreover, the probability
distributions of the fitted parameters are not assumed {\em a priori} to be Gaussian,
but are directly evaluated from the probability distributions  assigned to the experimental data;
\item  The possibility to automatically take into account   the  effects of the differential cross section systematics  in the error bars of the LEX coefficients of the DDPs.
\enitem

\noindent
The application of this strategy to the eight-parameter fit gave probability distributions with very broad and asymmetric tails in all cases,
except for $\alpha_{E1}$.  
For instance, the 68\% confidence interval of  $\alpha_{E1,\nu}$ 
was found to be  $\alpha_{E1,\nu} = (2.81^{+8.14}_{-6.44})\cdot 10^{-4}$ fm$^5$.

  This is a further confirmation that the quality of the present experimental database is too poor to allow  any meaningful estimate of the
  higher-order coefficients in the LEX of the scalar DDPs, considering 
      the very low sensitivity to them of the differential cross section below pion-production threshold.
      For this reason we reduced the number of free parameters to three: 
       the polarizability difference $\astat-\bstat$
       and the dispersive polarizabilities $\alpha_{E1,\nu}$ and $\beta_{M1,\nu}$~\cite{Pasquini:2007hf}. The remaining five parameters  were fixed 
       using the Baldin's sum rule value, smeared according to its resolution, for the polarizability sum $\alpha_{E1}+\beta_{M1}$, and DRs for the four parameters in the residual functions 
       $f_\alpha(\omega)$ and $f_\beta(\omega)$. In particular, the coefficients of the residual functions were fixed to describe the full energy-dependence of the scalar DDPs predicted from DRs (see Fig.~\ref{fig:alpha-beta}).

	\begin{table}
	\centering
	{\renewcommand\arraystretch{1.2}
\begin{tabular}{ccc}

\hline
	 & $\astat (10^{-4}\text{fm}^3)$ & $\bstat (10^{-4}\text{fm}^3)$\\
\hline

	fit 1& $11.8\pm0.2$ & $2.0\mp0.2$\\
	fit 2 & $11.9\pm0.2$ & $1.9\mp0.2$\\
	fit 3 & $11.8\pm0.2$ & $2.0\mp0.2$\\

	PDG & $11.2\pm0.4$ & $2.5\mp0.4$\\

\hline
\end{tabular}}
\caption{Results for the static polarizabilities, using different fitting procedures and the FULL data set. See text for further explanation.}
\label{tab:stat_check}
\end{table}	
\section{Results and discussions}
\label{sec:results}
In this section we discuss the results from  the new fitting strategy with three fit parameters.
 The results for the  LEX coefficients of the scalar DDPs in both the cases of  the FULL and TAPS data set are shown in \tref{tab:4p}.
The corresponding probability distributions are still Gaussian distributed, as  displayed in \fr{fig:all_prob_4p}. 

In the three-parameter fit, also the gradient method showed convergence (without the inclusion of systematic errors),
   but the covariance matrix was forced to be positive definite.
This feature casts doubts on the validity of the procedure even if the fitted parameter estimates turned out to be close to those given in \tref{tab:4p}.
	\begin{table}[h]
{\renewcommand\arraystretch{1.2}
\begin{tabular}{ccc}
\hline
	 & FULL & TAPS \\
\hline
$\astat \quad(10^{-4}\text{fm}^3)$ & $13.3\pm0.8$ & $11.6\pm1.1$ \\
$\alpha_{E1,\nu} \quad(10^{-4}\text{fm}^5)$ & $-8.8\pm2.5$ & $-3.2\pm3.1$ \\
$\bstat \quad(10^{-4}\text{fm}^3)$ & $0.4\mp0.9$ & $2.2\mp1.1$ \\
$\beta_{M1,\nu} \quad(10^{-4}\text{fm}^5)$ & $10.8\pm2.8$ & $5.1\pm3.7$\\
\hline
\end{tabular}}
\caption{Values of the LEX coefficients of the scalar DDPs from the fit to the FULL (second column) and TAPS (third column) data set in the three-parameter case.}
\label{tab:4p}
	\end{table}

\begin{figure}[b!]
\includegraphics[scale=0.85]{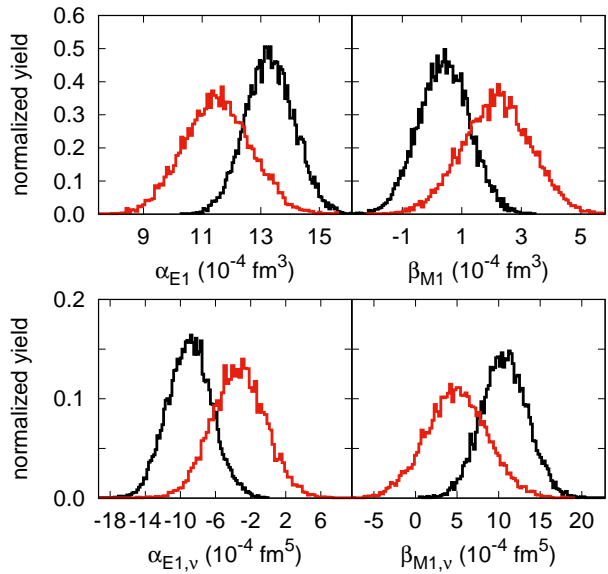}
\caption{Probability distributions of the fit parameters, with 100 bins per histogram,  from the bootstrap analysis of the FULL  (black lines)
and  TAPS data sets (red lines): static polarizabilities $\astat$ (top-left) and $\bstat$ (top-right) and dispersive polarizabilities
$\alpha_{E1,\nu}$ (bottom-left) and $\beta_{M1\nu}$ (bottom-right).}
\label{fig:all_prob_4p}
	\end{figure} 
Some comments are in order:
\beitem
\item It is likely that the FULL data set includes  inconsistent data, but the small number of experimental points in each data subset does not allow us to
  perform consistency checks without introducing biases;
\item The central values of the static dipole polarizabilities $\astat$ and $\bstat$ from the fit with FULL and the TAPS data sets
  are  different, but still compatible within the errors and in fairly good agreement with the PDG values.
  
This difference could also be  due to the
correlation between the different angular distributions of the two data sets and the varying sensitivity to  $\astat$ and $\bstat$
in different angular regions.
  
In~\fr{fig:distr}  we  show the two-dimensional joint probability
  distributions for $\astat,\bstat,\alpha_{E1,\nu}$ and $\beta_{M1,\nu}$.
  Apart from the strong correlation between $\astat$ and $\bstat$, which
  is due to the constraint of the Baldin's sum rule, 
  we observe significantly strong 
  correlations  also for all the other distributions.
  This is especially  true in the case of the dispersive
  polarizabilities $\alpha_{E,1\nu}$ and $\beta_{M1,\nu}$,
  that have very strong negative correlation  coefficients
  with $\astat$  and $\bstat$, respectively.

	\begin{figure}[t]
  \includegraphics[scale=.4]{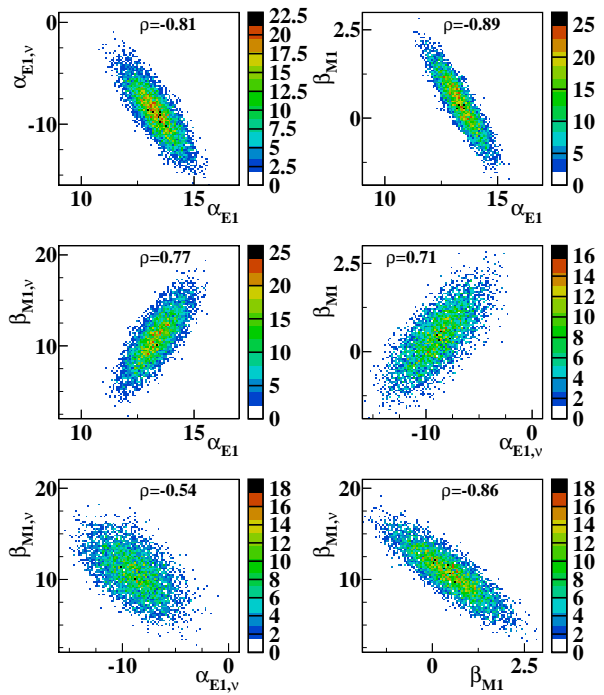}
  \caption{ Joint probability distributions and correlation coefficients
    $\rho$ for $\alpha_{E1}$, $\beta_{M1},$ $\alpha_{E1,\nu}$, and $\beta_{M1,\nu}$ for the  3-parameter fit of the FULL data base. The value ($\ne -1$) of the correlation coefficient between  $\alpha_{E1}$ and  $\beta_{M1}$, constrained by the Baldin's sum rule,  is due to the
uncertainty in the  sum rule value introduced in the bootstrap
procedure. The scales on the axes  are different for each plot.}
\label{fig:distr}
	\end{figure}

As already noticed before,     
this behavior is mainly a consequence of low sensitivity of the existing
data to the magnetic polarizabilities.

Even if this effect could also partially be due to inconsistent data subsets, as also recently discussed
  in~\cite{Krupina:2017pgr}, 
  only a relevant progress in both the quality and the quantity  of the existing data set 
  will allow to significantly improve this situation.   
  
If, as an example, we consider the data set defined in~\cite{Griesshammer:2012we,McGovern:2012ew}, we obtain for the fitted polarizabilities the following values: $\astat = (10.8 \pm 0.9) \cdot 10^{-4}\,\text{fm}^3$, $\alpha_{E,1\nu} =( -2.6 \pm 2.7) \cdot 10^{-4}\,\text{fm}^5$, $\bstat=( 2.9 \pm 1.0)\cdot 10^{-4}\,\text{fm}^3$ and $\beta_{M1,\nu} = (6.2 \pm 3.0)\cdot 10^{-4}\,\text{fm}^5$.
These values are compatible, within two standard deviations, with the ones given in ~\tref{tab:4p}.

 \enitem
Predictions  for the dispersive polarizabilities have been obtained within unsubtracted~\cite{Babusci:1998ww} and subtracted~\cite{Holstein:1999uu} DRs, and  baryon chiral perturbation theory~\cite{Lensky:2015awa}.
Our extraction of the dispersive polarizabilities is, within the large error bars, consistent with  these theoretical predictions and  
can not discriminate between them.
	\begin{figure}[t]
\includegraphics[scale=.64]{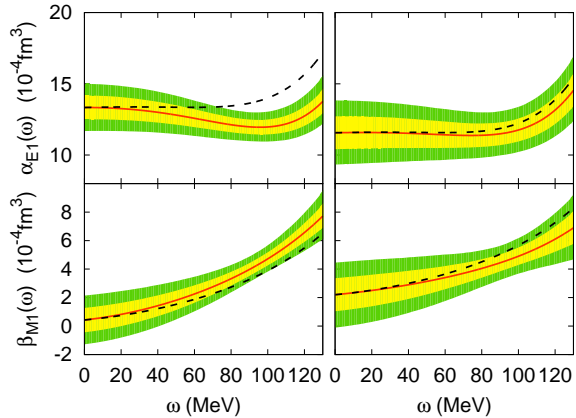}
\caption{Results from the fit of the scalar DDPs (solid line) for the FULL  (left panels) and TAPS (right panels) data sets: $\alpha_{E1}(\omega)$  on the top and $\beta_{M1}(\omega)$ on the bottom. The {$68\%$}
  (yellow)
  and {$95\%$}
  (green) {\em CL} areas 
  include all the correlations between the parameters. The dashed lines are the predictions from~DRs~\cite{Hildebrandt:2003fm}.}\label{fig:ddp_4p} 
	\end{figure}
        \noindent We pointed out that our fitting method provides a realistic probability distribution of the fitted parameters; another advantage is that it allows us to straightforwardly compute the error bands for the DDPs including all the correlations between the parameters.
In \fr{fig:ddp_4p}, we show our  fit results for the scalar DDPs, extracted from the FULL and TAPS data set,  as function of the cm energy $\omega$ and with the corresponding  $68\%$ and $95\%$ confidence level ({\em CL}) uncertainty bands.
They are compared with the subtracted DR predictions, obtained with the values of the static dipole polarizabilities from  the fit to the FULL and TAPS data sets in Tab.~\ref{tab:4p}.
The DR results for both the scalar DDPs are within the  $68\%$ confidence area of the fit results  for $\omega \lesssim 60$ MeV.
At higher energy,
the DR predictions for $\beta_{M1}(\omega)$ remain within  the $95\%$ {\em CL} region,   while
for $\alpha_{E1}(\omega)$ we observe deviations from the fit results in the case of the FULL data set and a very good agreement, within the $68\%$ confidence area,
in the case of  the TAPS data set. This different behavior can be again a hint of inconsistencies
between the two data sets.
The  larger relative error  in the case of  $\beta_{M1}(\omega)$ also reflects the lower sensitivity of the unpolarized RCS data to the magnetic polarizability than to the electric polarizability.

\section{Conclusions}
\label{sec:conclusion}

\noindent In summary, we have presented a new method based on the parametric bootstrap technique that allows us to extract, for the first time,
information  on the proton scalar DDPs from RCS data at low energies.
This method was never exploited so far to analyze Compton scattering data and has several advantages with respect to the standard $\chi^2$-minimization technique.
For example, it allows one to include in a straightforward way the systematic errors in the minimization procedure, without introducing a large number of additional fit parameters, and 
provides error distributions of the experimental data which are not assumed Gaussian {\it a priori}, but are directly evaluated from the probability distributions assigned to the data.

The extraction of the energy dependence of the DDPs turned out to be quite challenging, because  of the very low sensitivity of the unpolarized RCS
data to the higher-order dispersive coefficients. 
This gives  both large error bands of our estimates, in particular for $\beta_{M1}(\omega)$, and strong correlations between the
  fit parameters.
  
In the present analysis, the theoretical framework is based on dispersion relations, but it can be conveniently adapted to use other inputs, such as effective field theory calculations~\cite{Griesshammer:2012we,Lensky:2015awa} that have  been recently employed to extract the static polarizabilities.

Finally, future measurements  at MAMI~\cite{A2-Downie}
 hold the promise to improve the accuracy and the statistics of the data  and will help to determine with better accuracy the effects of the leading-order
static and dynamical polarizabilities. 

%


\begin{thebibliography}{48}%
\makeatletter
\providecommand \@ifxundefined [1]{%
 \@ifx{#1\undefined}
}%
\providecommand \@ifnum [1]{%
 \ifnum #1\expandafter \@firstoftwo
 \else \expandafter \@secondoftwo
 \fi
}%
\providecommand \@ifx [1]{%
 \ifx #1\expandafter \@firstoftwo
 \else \expandafter \@secondoftwo
 \fi
}%
\providecommand \natexlab [1]{#1}%
\providecommand \enquote  [1]{``#1''}%
\providecommand \bibnamefont  [1]{#1}%
\providecommand \bibfnamefont [1]{#1}%
\providecommand \citenamefont [1]{#1}%
\providecommand \href@noop [0]{\@secondoftwo}%
\providecommand \href [0]{\begingroup \@sanitize@url \@href}%
\providecommand \@href[1]{\@@startlink{#1}\@@href}%
\providecommand \@@href[1]{\endgroup#1\@@endlink}%
\providecommand \@sanitize@url [0]{\catcode `\\12\catcode `\$12\catcode
  `\&12\catcode `\#12\catcode `\^12\catcode `\_12\catcode `\%12\relax}%
\providecommand \@@startlink[1]{}%
\providecommand \@@endlink[0]{}%
\providecommand \url  [0]{\begingroup\@sanitize@url \@url }%
\providecommand \@url [1]{\endgroup\@href {#1}{\urlprefix }}%
\providecommand \urlprefix  [0]{URL }%
\providecommand \Eprint [0]{\href }%
\providecommand \doibase [0]{http://dx.doi.org/}%
\providecommand \selectlanguage [0]{\@gobble}%
\providecommand \bibinfo  [0]{\@secondoftwo}%
\providecommand \bibfield  [0]{\@secondoftwo}%
\providecommand \translation [1]{[#1]}%
\providecommand \BibitemOpen [0]{}%
\providecommand \bibitemStop [0]{}%
\providecommand \bibitemNoStop [0]{.\EOS\space}%
\providecommand \EOS [0]{\spacefactor3000\relax}%
\providecommand \BibitemShut  [1]{\csname bibitem#1\endcsname}%
\let\auto@bib@innerbib\@empty
\bibitem [{\citenamefont {Babusci}\ \emph {et~al.}(1998)\citenamefont
  {Babusci}, \citenamefont {Giordano}, \citenamefont {L'vov}, \citenamefont
  {Matone},\ and\ \citenamefont {Nathan}}]{Babusci:1998ww}%
  \BibitemOpen
  \bibfield  {author} {\bibinfo {author} {\bibfnamefont {D.}~\bibnamefont
  {Babusci}}, \bibinfo {author} {\bibfnamefont {G.}~\bibnamefont {Giordano}},
  \bibinfo {author} {\bibfnamefont {A.}~\bibnamefont {L'vov}}, \bibinfo
  {author} {\bibfnamefont {G.}~\bibnamefont {Matone}}, \ and\ \bibinfo {author}
  {\bibfnamefont {A.}~\bibnamefont {Nathan}},\ }\href {\doibase
  10.1103/PhysRevC.58.1013} {\bibfield  {journal} {\bibinfo  {journal} {Phys.
  Rev.}\ }\textbf {\bibinfo {volume} {C58}},\ \bibinfo {pages} {1013} (\bibinfo
  {year} {1998})},\ \Eprint {http://arxiv.org/abs/hep-ph/9803347}
  {arXiv:hep-ph/9803347 [hep-ph]} \BibitemShut {NoStop}%
\bibitem [{\citenamefont {Drechsel}\ \emph {et~al.}(2003)\citenamefont
  {Drechsel}, \citenamefont {Pasquini},\ and\ \citenamefont
  {Vanderhaeghen}}]{Drechsel:2002ar}%
  \BibitemOpen
  \bibfield  {author} {\bibinfo {author} {\bibfnamefont {D.}~\bibnamefont
  {Drechsel}}, \bibinfo {author} {\bibfnamefont {B.}~\bibnamefont {Pasquini}},
  \ and\ \bibinfo {author} {\bibfnamefont {M.}~\bibnamefont {Vanderhaeghen}},\
  }\href {\doibase 10.1016/S0370-1573(02)00636-1} {\bibfield  {journal}
  {\bibinfo  {journal} {Phys. Rept.}\ }\textbf {\bibinfo {volume} {378}},\
  \bibinfo {pages} {99} (\bibinfo {year} {2003})},\ \Eprint
  {http://arxiv.org/abs/hep-ph/0212124} {arXiv:hep-ph/0212124 [hep-ph]}
  \BibitemShut {NoStop}%
\bibitem [{\citenamefont {Schumacher}(2005)}]{Schumacher:2005an}%
  \BibitemOpen
  \bibfield  {author} {\bibinfo {author} {\bibfnamefont {M.}~\bibnamefont
  {Schumacher}},\ }\href {\doibase 10.1016/j.ppnp.2005.01.033} {\bibfield
  {journal} {\bibinfo  {journal} {Prog. Part. Nucl. Phys.}\ }\textbf {\bibinfo
  {volume} {55}},\ \bibinfo {pages} {567} (\bibinfo {year} {2005})},\ \Eprint
  {http://arxiv.org/abs/hep-ph/0501167} {arXiv:hep-ph/0501167 [hep-ph]}
  \BibitemShut {NoStop}%
\bibitem [{\citenamefont {Griesshammer}\ \emph {et~al.}(2012)\citenamefont
  {Griesshammer}, \citenamefont {McGovern}, \citenamefont {Phillips},\ and\
  \citenamefont {Feldman}}]{Griesshammer:2012we}%
  \BibitemOpen
  \bibfield  {author} {\bibinfo {author} {\bibfnamefont {H.~W.}\ \bibnamefont
  {Griesshammer}}, \bibinfo {author} {\bibfnamefont {J.~A.}\ \bibnamefont
  {McGovern}}, \bibinfo {author} {\bibfnamefont {D.~R.}\ \bibnamefont
  {Phillips}}, \ and\ \bibinfo {author} {\bibfnamefont {G.}~\bibnamefont
  {Feldman}},\ }\href {\doibase 10.1016/j.ppnp.2012.04.003} {\bibfield
  {journal} {\bibinfo  {journal} {Prog. Part. Nucl. Phys.}\ }\textbf {\bibinfo
  {volume} {67}},\ \bibinfo {pages} {841} (\bibinfo {year} {2012})},\ \Eprint
  {http://arxiv.org/abs/1203.6834} {arXiv:1203.6834 [nucl-th]} \BibitemShut
  {NoStop}%
\bibitem [{\citenamefont {Hagelstein}\ \emph {et~al.}(2016)\citenamefont
  {Hagelstein}, \citenamefont {Miskimen},\ and\ \citenamefont
  {Pascalutsa}}]{Hagelstein:2015egb}%
  \BibitemOpen
  \bibfield  {author} {\bibinfo {author} {\bibfnamefont {F.}~\bibnamefont
  {Hagelstein}}, \bibinfo {author} {\bibfnamefont {R.}~\bibnamefont
  {Miskimen}}, \ and\ \bibinfo {author} {\bibfnamefont {V.}~\bibnamefont
  {Pascalutsa}},\ }\href {\doibase 10.1016/j.ppnp.2015.12.001} {\bibfield
  {journal} {\bibinfo  {journal} {Prog. Part. Nucl. Phys.}\ }\textbf {\bibinfo
  {volume} {88}},\ \bibinfo {pages} {29} (\bibinfo {year} {2016})},\ \Eprint
  {http://arxiv.org/abs/1512.03765} {arXiv:1512.03765 [nucl-th]} \BibitemShut
  {NoStop}%
\bibitem [{\citenamefont {Pasquini}\ and\ \citenamefont
  {Vanderhaeghen}()}]{Pasquini:2018wbl}%
  \BibitemOpen
  \bibfield  {author} {\bibinfo {author} {\bibfnamefont {B.}~\bibnamefont
  {Pasquini}}\ and\ \bibinfo {author} {\bibfnamefont {M.}~\bibnamefont
  {Vanderhaeghen}},\ }\href@noop {} {\ }\Eprint
  {http://arxiv.org/abs/1805.10482} {arXiv:1805.10482 [hep-ph]} \BibitemShut
  {NoStop}%
\bibitem [{\citenamefont {Patrignani}\ \emph {et~al.}(2016)\citenamefont
  {Patrignani} \emph {et~al.}}]{Patrignani:2016xqp}%
  \BibitemOpen
  \bibfield  {author} {\bibinfo {author} {\bibfnamefont {C.}~\bibnamefont
  {Patrignani}} \emph {et~al.} (\bibinfo {collaboration} {Particle Data
  Group}),\ }\href {\doibase 10.1088/1674-1137/40/10/100001} {\bibfield
  {journal} {\bibinfo  {journal} {Chin. Phys.}\ }\textbf {\bibinfo {volume}
  {C40}},\ \bibinfo {pages} {100001} (\bibinfo {year} {2016})}\BibitemShut
  {NoStop}%
\bibitem [{\citenamefont {Martel}\ \emph {et~al.}(2015)\citenamefont {Martel}
  \emph {et~al.}}]{Martel:2014pba}%
  \BibitemOpen
  \bibfield  {author} {\bibinfo {author} {\bibfnamefont {P.~P.}\ \bibnamefont
  {Martel}} \emph {et~al.} (\bibinfo {collaboration} {A2}),\ }\href {\doibase
  10.1103/PhysRevLett.114.112501} {\bibfield  {journal} {\bibinfo  {journal}
  {Phys. Rev. Lett.}\ }\textbf {\bibinfo {volume} {114}},\ \bibinfo {pages}
  {112501} (\bibinfo {year} {2015})},\ \Eprint {http://arxiv.org/abs/1408.1576}
  {arXiv:1408.1576 [nucl-ex]} \BibitemShut {NoStop}%
\bibitem [{\citenamefont {Maize}\ and\ \citenamefont
  {Smetanka}(2008)}]{0143-0807-29-3-010}%
  \BibitemOpen
  \bibfield  {author} {\bibinfo {author} {\bibfnamefont {M.~A.}\ \bibnamefont
  {Maize}}\ and\ \bibinfo {author} {\bibfnamefont {J.~J.}\ \bibnamefont
  {Smetanka}},\ }\href {http://stacks.iop.org/0143-0807/29/i=3/a=010}
  {\bibfield  {journal} {\bibinfo  {journal} {European Journal of Physics}\
  }\textbf {\bibinfo {volume} {29}},\ \bibinfo {pages} {497} (\bibinfo {year}
  {2008})}\BibitemShut {NoStop}%
\bibitem [{\citenamefont {Kittel}(1996)}]{Kittel}%
  \BibitemOpen
  \bibfield  {author} {\bibinfo {author} {\bibfnamefont {C.}~\bibnamefont
  {Kittel}},\ }\href@noop {} {\emph {\bibinfo {title} {{Introduction to Solid
  State Physics}}}}\ (\bibinfo  {publisher} {Wiley, New York},\ \bibinfo {year}
  {1996})\BibitemShut {NoStop}%
\bibitem [{\citenamefont {{Glover}}\ and\ \citenamefont
  {{Weinhold}}(1977)}]{1977JChPh..66..191G}%
  \BibitemOpen
  \bibfield  {author} {\bibinfo {author} {\bibfnamefont {R.~M.}\ \bibnamefont
  {{Glover}}}\ and\ \bibinfo {author} {\bibfnamefont {F.}~\bibnamefont
  {{Weinhold}}},\ }\href {\doibase 10.1063/1.433653} {\bibfield  {journal}
  {\bibinfo  {journal} {\jcp}\ }\textbf {\bibinfo {volume} {66}},\ \bibinfo
  {pages} {191} (\bibinfo {year} {1977})}\BibitemShut {NoStop}%
\bibitem [{\citenamefont {Griesshammer}\ and\ \citenamefont
  {Hemmert}(2002)}]{Griesshammer:2001uw}%
  \BibitemOpen
  \bibfield  {author} {\bibinfo {author} {\bibfnamefont {H.~W.}\ \bibnamefont
  {Griesshammer}}\ and\ \bibinfo {author} {\bibfnamefont {T.~R.}\ \bibnamefont
  {Hemmert}},\ }\href {\doibase 10.1103/PhysRevC.65.045207} {\bibfield
  {journal} {\bibinfo  {journal} {Phys. Rev.}\ }\textbf {\bibinfo {volume}
  {C65}},\ \bibinfo {pages} {045207} (\bibinfo {year} {2002})},\ \Eprint
  {http://arxiv.org/abs/nucl-th/0110006} {arXiv:nucl-th/0110006 [nucl-th]}
  \BibitemShut {NoStop}%
\bibitem [{\citenamefont {Hildebrandt}\ \emph {et~al.}(2004)\citenamefont
  {Hildebrandt}, \citenamefont {Griesshammer}, \citenamefont {Hemmert},\ and\
  \citenamefont {Pasquini}}]{Hildebrandt:2003fm}%
  \BibitemOpen
  \bibfield  {author} {\bibinfo {author} {\bibfnamefont {R.~P.}\ \bibnamefont
  {Hildebrandt}}, \bibinfo {author} {\bibfnamefont {H.~W.}\ \bibnamefont
  {Griesshammer}}, \bibinfo {author} {\bibfnamefont {T.~R.}\ \bibnamefont
  {Hemmert}}, \ and\ \bibinfo {author} {\bibfnamefont {B.}~\bibnamefont
  {Pasquini}},\ }\href {\doibase 10.1140/epja/i2003-10144-9} {\bibfield
  {journal} {\bibinfo  {journal} {Eur. Phys. J.}\ }\textbf {\bibinfo {volume}
  {A20}},\ \bibinfo {pages} {293} (\bibinfo {year} {2004})},\ \Eprint
  {http://arxiv.org/abs/nucl-th/0307070} {arXiv:nucl-th/0307070 [nucl-th]}
  \BibitemShut {NoStop}%
\bibitem [{\citenamefont {Aleksejevs}\ and\ \citenamefont
  {Barkanova}(2013)}]{Aleksejevs:2013cda}%
  \BibitemOpen
  \bibfield  {author} {\bibinfo {author} {\bibfnamefont {A.}~\bibnamefont
  {Aleksejevs}}\ and\ \bibinfo {author} {\bibfnamefont {S.}~\bibnamefont
  {Barkanova}},\ }\href {\doibase 10.1016/j.nuclphysbps.2013.10.004} {\bibfield
   {journal} {\bibinfo  {journal} {Nucl. Phys. Proc. Suppl.}\ }\textbf
  {\bibinfo {volume} {245}},\ \bibinfo {pages} {17} (\bibinfo {year} {2013})},\
  \Eprint {http://arxiv.org/abs/1309.3313} {arXiv:1309.3313 [hep-ph]}
  \BibitemShut {NoStop}%
\bibitem [{\citenamefont {Aleksejevs}\ and\ \citenamefont
  {Barkanova}(2011)}]{Aleksejevs:2010zw}%
  \BibitemOpen
  \bibfield  {author} {\bibinfo {author} {\bibfnamefont {A.}~\bibnamefont
  {Aleksejevs}}\ and\ \bibinfo {author} {\bibfnamefont {S.}~\bibnamefont
  {Barkanova}},\ }\href {\doibase 10.1088/0954-3899/38/3/035004} {\bibfield
  {journal} {\bibinfo  {journal} {J. Phys.}\ }\textbf {\bibinfo {volume}
  {G38}},\ \bibinfo {pages} {035004} (\bibinfo {year} {2011})},\ \Eprint
  {http://arxiv.org/abs/1010.3457} {arXiv:1010.3457 [nucl-th]} \BibitemShut
  {NoStop}%
\bibitem [{\citenamefont {Lensky}\ \emph {et~al.}(2015)\citenamefont {Lensky},
  \citenamefont {McGovern},\ and\ \citenamefont {Pascalutsa}}]{Lensky:2015awa}%
  \BibitemOpen
  \bibfield  {author} {\bibinfo {author} {\bibfnamefont {V.}~\bibnamefont
  {Lensky}}, \bibinfo {author} {\bibfnamefont {J.}~\bibnamefont {McGovern}}, \
  and\ \bibinfo {author} {\bibfnamefont {V.}~\bibnamefont {Pascalutsa}},\
  }\href {\doibase 10.1140/epjc/s10052-015-3791-0} {\bibfield  {journal}
  {\bibinfo  {journal} {Eur. Phys. J.}\ }\textbf {\bibinfo {volume} {C75}},\
  \bibinfo {pages} {604} (\bibinfo {year} {2015})},\ \Eprint
  {http://arxiv.org/abs/1510.02794} {arXiv:1510.02794 [hep-ph]} \BibitemShut
  {NoStop}%
\bibitem [{\citenamefont {Davidson}\ and\ \citenamefont
  {Hinkley}(1997)}]{Davidson-Hinkley}%
  \BibitemOpen
  \bibfield  {author} {\bibinfo {author} {\bibfnamefont {A.~C.}\ \bibnamefont
  {Davidson}}\ and\ \bibinfo {author} {\bibfnamefont {D.~V.}\ \bibnamefont
  {Hinkley}},\ }\href@noop {} {\emph {\bibinfo {title} {Bootstrap Methods and
  their Application}}}\ (\bibinfo  {publisher} {Cambridge University Press},\
  \bibinfo {year} {1997})\BibitemShut {NoStop}%
\bibitem [{\citenamefont {Hildebrandt}(2005)}]{Hildebrandt:2005ix}%
  \BibitemOpen
  \bibfield  {author} {\bibinfo {author} {\bibfnamefont {R.~P.}\ \bibnamefont
  {Hildebrandt}},\ }\emph {\bibinfo {title} {{Elastic Compton Scattering from
  the Nucleon and Deuteron}}},\ \href@noop {} {Ph.D. thesis},\ \bibinfo
  {school} {Munich, Tech. U.} (\bibinfo {year} {2005}),\ \Eprint
  {http://arxiv.org/abs/nucl-th/0512064} {arXiv:nucl-th/0512064 [nucl-th]}
  \BibitemShut {NoStop}%
\bibitem [{\citenamefont {Griesshammer}(2005)}]{Griesshammer:2004yn}%
  \BibitemOpen
  \bibfield  {author} {\bibinfo {author} {\bibfnamefont {H.~W.}\ \bibnamefont
  {Griesshammer}},\ }\bibfield  {booktitle} {\emph {\bibinfo {booktitle}
  {{Lepton scattering and the structure of hadrons and nuclei. Proceedings,
  International School of nuclear physics, 26th Course, Erice, Italy, September
  16-24, 2004}}},\ }\href {\doibase 10.1016/j.ppnp.2005.01.009} {\bibfield
  {journal} {\bibinfo  {journal} {Prog. Part. Nucl. Phys.}\ }\textbf {\bibinfo
  {volume} {55}},\ \bibinfo {pages} {215} (\bibinfo {year} {2005})},\ \Eprint
  {http://arxiv.org/abs/nucl-th/0411080} {arXiv:nucl-th/0411080 [nucl-th]}
  \BibitemShut {NoStop}%
\bibitem [{\citenamefont {Krupina}\ \emph {et~al.}(2018)\citenamefont
  {Krupina}, \citenamefont {Lensky},\ and\ \citenamefont
  {Pascalutsa}}]{Krupina:2017pgr}%
  \BibitemOpen
  \bibfield  {author} {\bibinfo {author} {\bibfnamefont {N.}~\bibnamefont
  {Krupina}}, \bibinfo {author} {\bibfnamefont {V.}~\bibnamefont {Lensky}}, \
  and\ \bibinfo {author} {\bibfnamefont {V.}~\bibnamefont {Pascalutsa}},\
  }\href {\doibase 10.1016/j.physletb.2018.04.066} {\bibfield  {journal}
  {\bibinfo  {journal} {Phys. Lett.}\ }\textbf {\bibinfo {volume} {B782}},\
  \bibinfo {pages} {34} (\bibinfo {year} {2018})},\ \Eprint
  {http://arxiv.org/abs/1712.05349} {arXiv:1712.05349 [nucl-th]} \BibitemShut
  {NoStop}%
\bibitem [{\citenamefont {Lapidus}\ and\ \citenamefont
  {Chao}(1961)}]{Lapidus:1960}%
  \BibitemOpen
  \bibfield  {author} {\bibinfo {author} {\bibfnamefont {L.~I.}\ \bibnamefont
  {Lapidus}}\ and\ \bibinfo {author} {\bibfnamefont {C.~K.}\ \bibnamefont
  {Chao}},\ }\href@noop {} {\bibfield  {journal} {\bibinfo  {journal} {Sov.
  Phys. JETP}\ }\textbf {\bibinfo {volume} {14}},\ \bibinfo {pages} {210}
  (\bibinfo {year} {1961})}\BibitemShut {NoStop}%
\bibitem [{\citenamefont {Ritus}(1957)}]{Ritus}%
  \BibitemOpen
  \bibfield  {author} {\bibinfo {author} {\bibfnamefont {V.~I.}\ \bibnamefont
  {Ritus}},\ }\href@noop {} {\bibfield  {journal} {\bibinfo  {journal} {Sov.
  Phys. JETP}\ }\textbf {\bibinfo {volume} {5}},\ \bibinfo {pages} {1249}
  (\bibinfo {year} {1957})},\ \bibinfo {note} {[ZhETP
  32,1536(1957)]}\BibitemShut {NoStop}%
\bibitem [{\citenamefont {Contogouris}(1962)}]{Contogouris1962}%
  \BibitemOpen
  \bibfield  {author} {\bibinfo {author} {\bibfnamefont {A.~P.}\ \bibnamefont
  {Contogouris}},\ }\href {\doibase 10.1007/BF02733319} {\bibfield  {journal}
  {\bibinfo  {journal} {Il Nuovo Cimento}\ }\textbf {\bibinfo {volume} {25}},\
  \bibinfo {pages} {104} (\bibinfo {year} {1962})}\BibitemShut {NoStop}%
\bibitem [{\citenamefont {L'vov}\ \emph {et~al.}(1997)\citenamefont {L'vov},
  \citenamefont {Petrun'kin},\ and\ \citenamefont {Schumacher}}]{Lvov:1996rmi}%
  \BibitemOpen
  \bibfield  {author} {\bibinfo {author} {\bibfnamefont {A.~I.}\ \bibnamefont
  {L'vov}}, \bibinfo {author} {\bibfnamefont {V.~A.}\ \bibnamefont
  {Petrun'kin}}, \ and\ \bibinfo {author} {\bibfnamefont {M.}~\bibnamefont
  {Schumacher}},\ }\href {\doibase 10.1103/PhysRevC.55.359} {\bibfield
  {journal} {\bibinfo  {journal} {Phys. Rev.}\ }\textbf {\bibinfo {volume}
  {C55}},\ \bibinfo {pages} {359} (\bibinfo {year} {1997})}\BibitemShut
  {NoStop}%
\bibitem [{\citenamefont {Pasquini}\ \emph {et~al.}(2010)\citenamefont
  {Pasquini}, \citenamefont {Pedroni},\ and\ \citenamefont
  {Drechsel}}]{Pasquini:2010zr}%
  \BibitemOpen
  \bibfield  {author} {\bibinfo {author} {\bibfnamefont {B.}~\bibnamefont
  {Pasquini}}, \bibinfo {author} {\bibfnamefont {P.}~\bibnamefont {Pedroni}}, \
  and\ \bibinfo {author} {\bibfnamefont {D.}~\bibnamefont {Drechsel}},\ }\href
  {\doibase 10.1016/j.physletb.2010.03.007} {\bibfield  {journal} {\bibinfo
  {journal} {Phys. Lett.}\ }\textbf {\bibinfo {volume} {B687}},\ \bibinfo
  {pages} {160} (\bibinfo {year} {2010})},\ \Eprint
  {http://arxiv.org/abs/1001.4230} {arXiv:1001.4230 [hep-ph]} \BibitemShut
  {NoStop}%
\bibitem [{\citenamefont {Olmos~de Leon}\ \emph {et~al.}(2001)\citenamefont
  {Olmos~de Leon} \emph {et~al.}}]{OlmosdeLeon:2001zn}%
  \BibitemOpen
  \bibfield  {author} {\bibinfo {author} {\bibfnamefont {V.}~\bibnamefont
  {Olmos~de Leon}} \emph {et~al.},\ }\href {\doibase 10.1007/s100500170132}
  {\bibfield  {journal} {\bibinfo  {journal} {Eur. Phys. J.}\ }\textbf
  {\bibinfo {volume} {A10}},\ \bibinfo {pages} {207} (\bibinfo {year}
  {2001})}\BibitemShut {NoStop}%
\bibitem [{\citenamefont {Drechsel}\ \emph {et~al.}(1999)\citenamefont
  {Drechsel}, \citenamefont {Gorchtein}, \citenamefont {Pasquini},\ and\
  \citenamefont {Vanderhaeghen}}]{Drechsel:1999rf}%
  \BibitemOpen
  \bibfield  {author} {\bibinfo {author} {\bibfnamefont {D.}~\bibnamefont
  {Drechsel}}, \bibinfo {author} {\bibfnamefont {M.}~\bibnamefont {Gorchtein}},
  \bibinfo {author} {\bibfnamefont {B.}~\bibnamefont {Pasquini}}, \ and\
  \bibinfo {author} {\bibfnamefont {M.}~\bibnamefont {Vanderhaeghen}},\ }\href
  {\doibase 10.1103/PhysRevC.61.015204} {\bibfield  {journal} {\bibinfo
  {journal} {Phys. Rev.}\ }\textbf {\bibinfo {volume} {C61}},\ \bibinfo {pages}
  {015204} (\bibinfo {year} {1999})},\ \Eprint
  {http://arxiv.org/abs/hep-ph/9904290} {arXiv:hep-ph/9904290 [hep-ph]}
  \BibitemShut {NoStop}%
\bibitem [{\citenamefont {Drechsel}\ \emph {et~al.}(2007)\citenamefont
  {Drechsel}, \citenamefont {Kamalov},\ and\ \citenamefont
  {Tiator}}]{Drechsel:2007if}%
  \BibitemOpen
  \bibfield  {author} {\bibinfo {author} {\bibfnamefont {D.}~\bibnamefont
  {Drechsel}}, \bibinfo {author} {\bibfnamefont {S.~S.}\ \bibnamefont
  {Kamalov}}, \ and\ \bibinfo {author} {\bibfnamefont {L.}~\bibnamefont
  {Tiator}},\ }\href {\doibase 10.1140/epja/i2007-10490-6} {\bibfield
  {journal} {\bibinfo  {journal} {Eur. Phys. J.}\ }\textbf {\bibinfo {volume}
  {A34}},\ \bibinfo {pages} {69} (\bibinfo {year} {2007})},\ \Eprint
  {http://arxiv.org/abs/0710.0306} {arXiv:0710.0306 [nucl-th]} \BibitemShut
  {NoStop}%
\bibitem [{\citenamefont {Pasquini}\ \emph {et~al.}(2007)\citenamefont
  {Pasquini}, \citenamefont {Drechsel},\ and\ \citenamefont
  {Vanderhaeghen}}]{Pasquini:2007hf}%
  \BibitemOpen
  \bibfield  {author} {\bibinfo {author} {\bibfnamefont {B.}~\bibnamefont
  {Pasquini}}, \bibinfo {author} {\bibfnamefont {D.}~\bibnamefont {Drechsel}},
  \ and\ \bibinfo {author} {\bibfnamefont {M.}~\bibnamefont {Vanderhaeghen}},\
  }\href {\doibase 10.1103/PhysRevC.76.015203} {\bibfield  {journal} {\bibinfo
  {journal} {Phys. Rev.}\ }\textbf {\bibinfo {volume} {C76}},\ \bibinfo {pages}
  {015203} (\bibinfo {year} {2007})},\ \Eprint {http://arxiv.org/abs/0705.0282}
  {arXiv:0705.0282 [hep-ph]} \BibitemShut {NoStop}%
\bibitem [{\citenamefont {Holstein}\ \emph {et~al.}(2000)\citenamefont
  {Holstein}, \citenamefont {Drechsel}, \citenamefont {Pasquini},\ and\
  \citenamefont {Vanderhaeghen}}]{Holstein:1999uu}%
  \BibitemOpen
  \bibfield  {author} {\bibinfo {author} {\bibfnamefont {B.~R.}\ \bibnamefont
  {Holstein}}, \bibinfo {author} {\bibfnamefont {D.}~\bibnamefont {Drechsel}},
  \bibinfo {author} {\bibfnamefont {B.}~\bibnamefont {Pasquini}}, \ and\
  \bibinfo {author} {\bibfnamefont {M.}~\bibnamefont {Vanderhaeghen}},\ }\href
  {\doibase 10.1103/PhysRevC.61.034316} {\bibfield  {journal} {\bibinfo
  {journal} {Phys. Rev.}\ }\textbf {\bibinfo {volume} {C61}},\ \bibinfo {pages}
  {034316} (\bibinfo {year} {2000})},\ \Eprint
  {http://arxiv.org/abs/hep-ph/9910427} {arXiv:hep-ph/9910427 [hep-ph]}
  \BibitemShut {NoStop}%
\bibitem [{\citenamefont {Oxley}(1958)}]{Oxley:1958zz}%
  \BibitemOpen
  \bibfield  {author} {\bibinfo {author} {\bibfnamefont {C.~L.}\ \bibnamefont
  {Oxley}},\ }\href {\doibase 10.1103/PhysRev.110.733} {\bibfield  {journal}
  {\bibinfo  {journal} {Phys. Rev.}\ }\textbf {\bibinfo {volume} {110}},\
  \bibinfo {pages} {733} (\bibinfo {year} {1958})}\BibitemShut {NoStop}%
\bibitem [{\citenamefont {Hyman}\ \emph {et~al.}(1959)\citenamefont {Hyman},
  \citenamefont {Ely}, \citenamefont {Frisch},\ and\ \citenamefont
  {Wahlig}}]{Hyman:1959zz}%
  \BibitemOpen
  \bibfield  {author} {\bibinfo {author} {\bibfnamefont {L.~G.}\ \bibnamefont
  {Hyman}}, \bibinfo {author} {\bibfnamefont {R.}~\bibnamefont {Ely}}, \bibinfo
  {author} {\bibfnamefont {D.~H.}\ \bibnamefont {Frisch}}, \ and\ \bibinfo
  {author} {\bibfnamefont {M.~A.}\ \bibnamefont {Wahlig}},\ }\href {\doibase
  10.1103/PhysRevLett.3.93} {\bibfield  {journal} {\bibinfo  {journal} {Phys.
  Rev. Lett.}\ }\textbf {\bibinfo {volume} {3}},\ \bibinfo {pages} {93}
  (\bibinfo {year} {1959})}\BibitemShut {NoStop}%
\bibitem [{\citenamefont {Goldansky}\ \emph {et~al.}(1960)\citenamefont
  {Goldansky}, \citenamefont {Karpukhin}, \citenamefont {Kutsenko},\ and\
  \citenamefont {Pavlovskaya}}]{GOLDANSKY1960473}%
  \BibitemOpen
  \bibfield  {author} {\bibinfo {author} {\bibfnamefont {V.}~\bibnamefont
  {Goldansky}}, \bibinfo {author} {\bibfnamefont {O.}~\bibnamefont
  {Karpukhin}}, \bibinfo {author} {\bibfnamefont {A.}~\bibnamefont {Kutsenko}},
  \ and\ \bibinfo {author} {\bibfnamefont {V.}~\bibnamefont {Pavlovskaya}},\
  }\href {\doibase https://doi.org/10.1016/0029-5582(60)90418-1} {\bibfield
  {journal} {\bibinfo  {journal} {Nuclear Physics}\ }\textbf {\bibinfo {volume}
  {18}},\ \bibinfo {pages} {473 } (\bibinfo {year} {1960})}\BibitemShut
  {NoStop}%
\bibitem [{\citenamefont {Pugh}\ \emph {et~al.}(1957)\citenamefont {Pugh},
  \citenamefont {Gomez}, \citenamefont {Frisch},\ and\ \citenamefont
  {Janes}}]{Pugh:1957zz}%
  \BibitemOpen
  \bibfield  {author} {\bibinfo {author} {\bibfnamefont {G.~E.}\ \bibnamefont
  {Pugh}}, \bibinfo {author} {\bibfnamefont {R.}~\bibnamefont {Gomez}},
  \bibinfo {author} {\bibfnamefont {D.~H.}\ \bibnamefont {Frisch}}, \ and\
  \bibinfo {author} {\bibfnamefont {G.~S.}\ \bibnamefont {Janes}},\ }\href
  {\doibase 10.1103/PhysRev.105.982} {\bibfield  {journal} {\bibinfo  {journal}
  {Phys. Rev.}\ }\textbf {\bibinfo {volume} {105}},\ \bibinfo {pages} {982}
  (\bibinfo {year} {1957})}\BibitemShut {NoStop}%
\bibitem [{\citenamefont {Baranov}\ \emph {et~al.}(2001)\citenamefont
  {Baranov}, \citenamefont {L'vov}, \citenamefont {Petrunkin},\ and\
  \citenamefont {Shtarkov}}]{Baranov:2001jv}%
  \BibitemOpen
  \bibfield  {author} {\bibinfo {author} {\bibfnamefont {P.~S.}\ \bibnamefont
  {Baranov}}, \bibinfo {author} {\bibfnamefont {A.~I.}\ \bibnamefont {L'vov}},
  \bibinfo {author} {\bibfnamefont {V.~A.}\ \bibnamefont {Petrunkin}}, \ and\
  \bibinfo {author} {\bibfnamefont {L.~N.}\ \bibnamefont {Shtarkov}},\
  }\href@noop {} {\bibfield  {journal} {\bibinfo  {journal} {Phys. Part.
  Nucl.}\ }\textbf {\bibinfo {volume} {32}},\ \bibinfo {pages} {376} (\bibinfo
  {year} {2001})},\ \bibinfo {note} {[Fiz. Elem. Chast. Atom.
  Yadra32,699(2001)]}\BibitemShut {NoStop}%
\bibitem [{\citenamefont {Federspiel}\ \emph {et~al.}(1991)\citenamefont
  {Federspiel}, \citenamefont {Eisenstein}, \citenamefont {Lucas},
  \citenamefont {MacGibbon}, \citenamefont {Mellendorf}, \citenamefont
  {Nathan}, \citenamefont {O'Neill},\ and\ \citenamefont
  {Wells}}]{Federspiel:1991yd}%
  \BibitemOpen
  \bibfield  {author} {\bibinfo {author} {\bibfnamefont {F.~J.}\ \bibnamefont
  {Federspiel}}, \bibinfo {author} {\bibfnamefont {R.~A.}\ \bibnamefont
  {Eisenstein}}, \bibinfo {author} {\bibfnamefont {M.~A.}\ \bibnamefont
  {Lucas}}, \bibinfo {author} {\bibfnamefont {B.~E.}\ \bibnamefont
  {MacGibbon}}, \bibinfo {author} {\bibfnamefont {K.}~\bibnamefont
  {Mellendorf}}, \bibinfo {author} {\bibfnamefont {A.~M.}\ \bibnamefont
  {Nathan}}, \bibinfo {author} {\bibfnamefont {A.}~\bibnamefont {O'Neill}}, \
  and\ \bibinfo {author} {\bibfnamefont {D.~P.}\ \bibnamefont {Wells}},\ }\href
  {\doibase 10.1103/PhysRevLett.67.1511} {\bibfield  {journal} {\bibinfo
  {journal} {Phys. Rev. Lett.}\ }\textbf {\bibinfo {volume} {67}},\ \bibinfo
  {pages} {1511} (\bibinfo {year} {1991})}\BibitemShut {NoStop}%
\bibitem [{\citenamefont {Zieger}\ \emph {et~al.}(1992)\citenamefont {Zieger},
  \citenamefont {Van~de Vyver}, \citenamefont {Christmann}, \citenamefont
  {De~Graeve}, \citenamefont {Van~den Abeele},\ and\ \citenamefont
  {Ziegler}}]{Zieger:1992jq}%
  \BibitemOpen
  \bibfield  {author} {\bibinfo {author} {\bibfnamefont {A.}~\bibnamefont
  {Zieger}}, \bibinfo {author} {\bibfnamefont {R.}~\bibnamefont {Van~de
  Vyver}}, \bibinfo {author} {\bibfnamefont {D.}~\bibnamefont {Christmann}},
  \bibinfo {author} {\bibfnamefont {A.}~\bibnamefont {De~Graeve}}, \bibinfo
  {author} {\bibfnamefont {C.}~\bibnamefont {Van~den Abeele}}, \ and\ \bibinfo
  {author} {\bibfnamefont {B.}~\bibnamefont {Ziegler}},\ }\href {\doibase
  10.1016/0370-2693(92)90707-B} {\bibfield  {journal} {\bibinfo  {journal}
  {Phys. Lett.}\ }\textbf {\bibinfo {volume} {B278}},\ \bibinfo {pages} {34}
  (\bibinfo {year} {1992})}\BibitemShut {NoStop}%
\bibitem [{\citenamefont {MacGibbon}\ \emph {et~al.}(1995)\citenamefont
  {MacGibbon}, \citenamefont {Garino}, \citenamefont {Lucas}, \citenamefont
  {Nathan}, \citenamefont {Feldman},\ and\ \citenamefont
  {Dolbilkin}}]{MacGibbon:1995in}%
  \BibitemOpen
  \bibfield  {author} {\bibinfo {author} {\bibfnamefont {B.~E.}\ \bibnamefont
  {MacGibbon}}, \bibinfo {author} {\bibfnamefont {G.}~\bibnamefont {Garino}},
  \bibinfo {author} {\bibfnamefont {M.~A.}\ \bibnamefont {Lucas}}, \bibinfo
  {author} {\bibfnamefont {A.~M.}\ \bibnamefont {Nathan}}, \bibinfo {author}
  {\bibfnamefont {G.}~\bibnamefont {Feldman}}, \ and\ \bibinfo {author}
  {\bibfnamefont {B.}~\bibnamefont {Dolbilkin}},\ }\href {\doibase
  10.1103/PhysRevC.52.2097} {\bibfield  {journal} {\bibinfo  {journal} {Phys.
  Rev.}\ }\textbf {\bibinfo {volume} {C52}},\ \bibinfo {pages} {2097} (\bibinfo
  {year} {1995})},\ \Eprint {http://arxiv.org/abs/nucl-ex/9507001}
  {arXiv:nucl-ex/9507001 [nucl-ex]} \BibitemShut {NoStop}%
\bibitem [{\citenamefont {Hallin}\ \emph {et~al.}(1993)\citenamefont {Hallin}
  \emph {et~al.}}]{Hallin:1993ft}%
  \BibitemOpen
  \bibfield  {author} {\bibinfo {author} {\bibfnamefont {E.~L.}\ \bibnamefont
  {Hallin}} \emph {et~al.},\ }\href {\doibase 10.1103/PhysRevC.48.1497}
  {\bibfield  {journal} {\bibinfo  {journal} {Phys. Rev.}\ }\textbf {\bibinfo
  {volume} {C48}},\ \bibinfo {pages} {1497} (\bibinfo {year}
  {1993})}\BibitemShut {NoStop}%
\bibitem [{\citenamefont {Bernardini}\ \emph {et~al.}(1960)\citenamefont
  {Bernardini}, \citenamefont {Hanson}, \citenamefont {Odian}, \citenamefont
  {Yamagata}, \citenamefont {Auerbach},\ and\ \citenamefont
  {Filosofo}}]{Bernardini1960}%
  \BibitemOpen
  \bibfield  {author} {\bibinfo {author} {\bibfnamefont {G.}~\bibnamefont
  {Bernardini}}, \bibinfo {author} {\bibfnamefont {A.~O.}\ \bibnamefont
  {Hanson}}, \bibinfo {author} {\bibfnamefont {A.~C.}\ \bibnamefont {Odian}},
  \bibinfo {author} {\bibfnamefont {T.}~\bibnamefont {Yamagata}}, \bibinfo
  {author} {\bibfnamefont {L.~B.}\ \bibnamefont {Auerbach}}, \ and\ \bibinfo
  {author} {\bibfnamefont {I.}~\bibnamefont {Filosofo}},\ }\href {\doibase
  10.1007/BF02733177} {\bibfield  {journal} {\bibinfo  {journal} {Il Nuovo
  Cimento (1955-1965)}\ }\textbf {\bibinfo {volume} {18}},\ \bibinfo {pages}
  {1203} (\bibinfo {year} {1960})}\BibitemShut {NoStop}%
\bibitem [{\citenamefont {Baranov}\ \emph {et~al.}(1974)\citenamefont
  {Baranov}, \citenamefont {Buinov}, \citenamefont {Godin}, \citenamefont
  {Kuznetzova}, \citenamefont {Petrunkin}, \citenamefont {Tatarinskaya},
  \citenamefont {Shirthenko}, \citenamefont {Shtarkov}, \citenamefont
  {Yurtchenko},\ and\ \citenamefont {Yanulis}}]{Baranov:1974ec}%
  \BibitemOpen
  \bibfield  {author} {\bibinfo {author} {\bibfnamefont {P.}~\bibnamefont
  {Baranov}}, \bibinfo {author} {\bibfnamefont {G.}~\bibnamefont {Buinov}},
  \bibinfo {author} {\bibfnamefont {V.}~\bibnamefont {Godin}}, \bibinfo
  {author} {\bibfnamefont {V.}~\bibnamefont {Kuznetzova}}, \bibinfo {author}
  {\bibfnamefont {V.}~\bibnamefont {Petrunkin}}, \bibinfo {author}
  {\bibnamefont {Tatarinskaya}}, \bibinfo {author} {\bibfnamefont
  {V.}~\bibnamefont {Shirthenko}}, \bibinfo {author} {\bibfnamefont
  {L.}~\bibnamefont {Shtarkov}}, \bibinfo {author} {\bibfnamefont
  {V.}~\bibnamefont {Yurtchenko}}, \ and\ \bibinfo {author} {\bibfnamefont
  {{\relax Yu}.}~\bibnamefont {Yanulis}},\ }\href {\doibase
  10.1016/0370-2693(74)90736-9} {\bibfield  {journal} {\bibinfo  {journal}
  {Phys. Lett.}\ }\textbf {\bibinfo {volume} {52B}},\ \bibinfo {pages} {122}
  (\bibinfo {year} {1974})}\BibitemShut {NoStop}%
\bibitem [{\citenamefont {Baranov}\ \emph {et~al.}(1975)\citenamefont
  {Baranov}, \citenamefont {Buinov}, \citenamefont {Godin}, \citenamefont
  {Kuznetsova}, \citenamefont {Petrunkin}, \citenamefont {Tatarinskaya},
  \citenamefont {Shirchenko}, \citenamefont {Shtarkov}, \citenamefont
  {Yurchenko},\ and\ \citenamefont {Yanulis}}]{Baranov:1975ju}%
  \BibitemOpen
  \bibfield  {author} {\bibinfo {author} {\bibfnamefont {P.~S.}\ \bibnamefont
  {Baranov}}, \bibinfo {author} {\bibfnamefont {G.~M.}\ \bibnamefont {Buinov}},
  \bibinfo {author} {\bibfnamefont {V.~G.}\ \bibnamefont {Godin}}, \bibinfo
  {author} {\bibfnamefont {V.~A.}\ \bibnamefont {Kuznetsova}}, \bibinfo
  {author} {\bibfnamefont {V.~A.}\ \bibnamefont {Petrunkin}}, \bibinfo {author}
  {\bibfnamefont {L.~S.}\ \bibnamefont {Tatarinskaya}}, \bibinfo {author}
  {\bibfnamefont {V.~S.}\ \bibnamefont {Shirchenko}}, \bibinfo {author}
  {\bibfnamefont {L.~N.}\ \bibnamefont {Shtarkov}}, \bibinfo {author}
  {\bibfnamefont {V.~V.}\ \bibnamefont {Yurchenko}}, \ and\ \bibinfo {author}
  {\bibfnamefont {{\relax Yu}.~P.}\ \bibnamefont {Yanulis}},\ }\href@noop {}
  {\bibfield  {journal} {\bibinfo  {journal} {Yad. Fiz.}\ }\textbf {\bibinfo
  {volume} {21}},\ \bibinfo {pages} {689} (\bibinfo {year} {1975})}\BibitemShut
  {NoStop}%
\bibitem [{\citenamefont {McGovern}\ \emph {et~al.}(2013)\citenamefont
  {McGovern}, \citenamefont {Phillips},\ and\ \citenamefont
  {Griesshammer}}]{McGovern:2012ew}%
  \BibitemOpen
  \bibfield  {author} {\bibinfo {author} {\bibfnamefont {J.~A.}\ \bibnamefont
  {McGovern}}, \bibinfo {author} {\bibfnamefont {D.~R.}\ \bibnamefont
  {Phillips}}, \ and\ \bibinfo {author} {\bibfnamefont {H.~W.}\ \bibnamefont
  {Griesshammer}},\ }\href {\doibase 10.1140/epja/i2013-13012-1} {\bibfield
  {journal} {\bibinfo  {journal} {Eur. Phys. J.}\ }\textbf {\bibinfo {volume}
  {A49}},\ \bibinfo {pages} {12} (\bibinfo {year} {2013})},\ \Eprint
  {http://arxiv.org/abs/1210.4104} {arXiv:1210.4104 [nucl-th]} \BibitemShut
  {NoStop}%
\bibitem [{\citenamefont {Pasquini}\ \emph {et~al.}()\citenamefont {Pasquini},
  \citenamefont {Pedroni},\ and\ \citenamefont {Sconfietti}}]{preparation}%
  \BibitemOpen
  \bibfield  {author} {\bibinfo {author} {\bibfnamefont {B.}~\bibnamefont
  {Pasquini}}, \bibinfo {author} {\bibfnamefont {P.}~\bibnamefont {Pedroni}}, \
  and\ \bibinfo {author} {\bibfnamefont {S.}~\bibnamefont {Sconfietti}},\
  }\href@noop {} {\bibinfo  {journal} {in preparation}\ }\BibitemShut {NoStop}%
\bibitem [{\citenamefont {James}\ and\ \citenamefont
  {Roos}(1975)}]{James:1975dr}%
  \BibitemOpen
\bibfield  {journal} {  }\bibfield  {author} {\bibinfo {author} {\bibfnamefont
  {F.}~\bibnamefont {James}}\ and\ \bibinfo {author} {\bibfnamefont
  {M.}~\bibnamefont {Roos}},\ }\href {\doibase 10.1016/0010-4655(75)90039-9}
  {\bibfield  {journal} {\bibinfo  {journal} {Comput. Phys. Commun.}\ }\textbf
  {\bibinfo {volume} {10}},\ \bibinfo {pages} {343} (\bibinfo {year}
  {1975})}\BibitemShut {NoStop}%
\bibitem [{\citenamefont {Nelder}\ and\ \citenamefont
  {Mead}(1965)}]{Nelder:1965zz}%
  \BibitemOpen
  \bibfield  {author} {\bibinfo {author} {\bibfnamefont {J.~A.}\ \bibnamefont
  {Nelder}}\ and\ \bibinfo {author} {\bibfnamefont {R.}~\bibnamefont {Mead}},\
  }\href {\doibase 10.1093/comjnl/7.4.308} {\bibfield  {journal} {\bibinfo
  {journal} {Comput. J.}\ }\textbf {\bibinfo {volume} {7}},\ \bibinfo {pages}
  {308} (\bibinfo {year} {1965})}\BibitemShut {NoStop}%
\bibitem [{\citenamefont {James}(2006)}]{James:2006zz}%
  \BibitemOpen
  \bibfield  {author} {\bibinfo {author} {\bibfnamefont {F.}~\bibnamefont
  {James}},\ }\href@noop {} {\emph {\bibinfo {title} {{Statistical methods in
  experimental physics}}}}\ (\bibinfo  {publisher} {Hackensack, USA: World
  Scientific},\ \bibinfo {year} {2006})\BibitemShut {NoStop}%
\bibitem [{\citenamefont {Downie}\ and\ \citenamefont
  {et~al.}(2016)}]{A2-Downie}%
  \BibitemOpen
  \bibfield  {author} {\bibinfo {author} {\bibfnamefont {E.~J.}\ \bibnamefont
  {Downie}}\ and\ \bibinfo {author} {\bibnamefont {et~al.}},\ }\href@noop {}
  {\bibfield  {journal} {\bibinfo  {journal} {Proposal MAMI-A2/04-16}\ }
  (\bibinfo {year} {2016})}\BibitemShut {NoStop}%
\end{thebibliography}
\end{document}